\documentclass[aps,pre,reprint,showpacs,floatfix]{revtex4-1}
\usepackage{graphicx}
\usepackage{epstopdf}
\usepackage{epsfig}
\usepackage{natbib}
\begin{document}
%\title{Bidirectional Transport of Motor-driven Cargoes is a Random Walk with Memory}
\title{Memory, Bias and Correlations in Bidirectional Transport of Molecular Motor-driven Cargoes}
\author{Deepak Bhat}
\email{deepak@physics.iitm.ac.in}
\author{Manoj Gopalakrishnan}
\email{manoj@physics.iitm.ac.in}
\affiliation{Department of Physics, Indian Institute of Technology Madras, Chennai 600036, India}
\date{\today}
\begin{abstract}
Molecular motors are specialized proteins which perform active, directed transport of cellular cargoes on cytoskeletal filaments. In many cases, cargo 
motion powered by motor proteins is found to be bidirectional, and may be viewed as a biased random walk with fast unidirectional runs interspersed 
with slow `tug-of-war' states. The statistical properties of this walk are not known in detail, and here, we study memory and bias, as well as  directional correlations between successive runs in bidirectional transport. 
We show, based on a study of the direction reversal probabilities of the cargo using a purely stochastic (tug-of-war) model, that bidirectional motion 
of cellular cargoes is, in general, a correlated random walk. In particular, while the motion of a cargo driven by two oppositely pulling motors is a 
Markovian random walk,  memory of direction appears when multiple motors haul the cargo in one or both directions. In the latter case, the Markovian 
nature of the underlying single motor processes is hidden by internal transitions between degenerate run and pause states of the cargo. Interestingly, 
memory is found to be a non-monotonic function of the number of motors. Stochastic numerical simulations of the tug-of-war model support our 
mathematical results and extend them to biologically relevant situations. 
\end{abstract}
\maketitle 

\section{Introduction} 

Motor proteins are enzymes that convert chemical energy derived from hydrolysis of adenosine tri-phosphate (ATP) to mechanical work. Dynein and 
kinesin are two such proteins which perform directed motion on microtubules, in opposite directions. While a complete understanding of the process 
remains an open question, various plausible mechanisms leading to the directed transport have been discussed in the literature\cite{Astumian,Astumian2,Fisher,Fisher2,Magnasco1,Prost,Parmeggiani,Julicher,Chowdhury,Fisher3,Zhang,Debashish}.
Motor-driven cargo transport on cytoskeletal network interests biologists and physicists alike because of its relevance in understanding spatial 
organization of various organelles inside eukaryotic cells and because of the opportunities it provides for detailed quantitative modeling\cite{Badoual,Klumpp,SKlumpp,Muller,Mjmuller}. 
Although the primary purpose of molecular motors would appear to be fast unidirectional transport, many motor driven cargoes on microtubule filaments 
are found to move in bidirectional fashion\cite{Gross,Welte}. While tug-of-war (TOW) model explains bidirectional transport as a natural consequence 
of motors of opposite polarity (eg., kinesin and dynein) being simultaneously active and exerting forces on the cargo\cite{Muller,Gross,Welte}, 
regulated coordination model presumes the presence of a coordinating complex in the cargo which permits only one set of motors to be active at any 
point of time. 

\begin{figure}
\begin{center}
\begin{minipage}{3.85cm}
\includegraphics[width=3.9cm,height=3.3cm]{fig1a.eps}
% staterepresent.ps: 595x842 pixel, 72dpi, 20.99x29.70 cm, bb=0 0 595 842
\end{minipage}
\begin{minipage}{4cm}
\includegraphics[width=3.2cm,keepaspectratio=true,angle=-90]{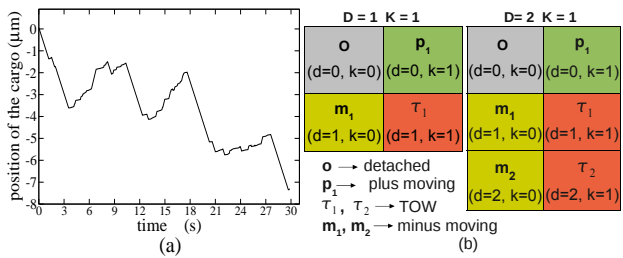}
% staterepresent.ps: 595x842 pixel, 72dpi, 20.99x29.70 cm, bb=0 0 595 842
\end{minipage}
\caption{\label{fig:fig1} (color online) (a) A typical trajectory of a bidirectionally moving cargo, as generated in our simulations (see \cite{Deepak} for details). 
(b) State representation of a cargo driven by a dynein and a kinesin, two dyneins and a kinesin. Here, motility states of the cargo are non-degenerate 
in the first case, but degenerate in the second.}
\end{center}
\end{figure}

A typical bidirectional cargo is hauled by several motors of opposite directionality, and would have a definite drift towards the plus or minus end 
of the filament. A characteristic trajectory of bidirectional cargo hauled by five dyneins and a kinesin, generated in our simulations is shown in 
Fig. \ref{fig:fig1}(a)  (see ref.\cite{Deepak} for details). Such a motion may be visualized as a biased random walk \cite{Maly}, with unidirectional 
runs separated by pause states. In a recent in vitro experimental study, presence of memory in bidirectional motion was observed \cite{Leidel},  
wherein a bidirectional cargo stalled by an optical trap while moving was found to move preferentially in the pre-stall direction after detachment 
from the filament under the influence of the trap. While a detailed study of this experiment is outside the scope of the present paper, it is 
pertinent to ask: is motor protein-powered bidirectional organelle transport a Markovian random walk?  

In the present paper, we study the history dependence of bidirectional cargo motion powered by molecular motors within the framework of the stochastic 
TOW model. Our rigorous mathematical calculations, supported by stochastic simulations, show that bidirectional motor-mediated transport is a 
non-Markovian random walk, characterized by multi-exponential waiting time distributions. Interestingly, a recent theoretical work \cite{Yunxin} has 
studied the effects of pre-assigned memory in  transition rates on bidirectional cargo transport, but does not discuss its origins. By contrast, our 
work shows explicitly how memory emerges as a consequence of the degeneracy of the states of motion of the cargo, when hauled by multiple motor teams. 
We also find that correlation between run directions is likely to extend to several TOW events in typical experimental situations.

Several examples of persistent (correlated) random walks are known in biology, e.g., bacterial chemotaxis\cite{schnitzer} and locomotion of 
slime mold amoeba {\it D. discoideum}\cite{sambeth}. However, typically in such cases, the underlying mechanism behind persistence of direction is 
not precisely known. Bidirectional cargo transport by molecular motors, on the other hand, can be reconstituted in vitro and the number of cargo-bound 
motors estimated using optical trap; therefore, it is a much more controllable system compared to the previous examples. Further, the underlying 
fundamental processes (binding and unbinding of individual motors) are Markovian, and therefore, the present system constitutes a fine illustration 
of how apparent non-Markovian behavior emerges from purely Markovian state transitions underneath. 

\section{Model and Formalism}
The stochastic TOW model \cite{Muller}, assumes a fixed number of minus-moving dyneins ($D$) and plus-moving kinesins ($K$) on a cargo. Each motor on 
the cargo binds to and unbinds from the microtubule stochastically with rates $\pi_{\pm}$ and $\epsilon_{\pm}$ respectively, with $+$ subscript for 
kinesin and $-$ for dynein, while it is assumed that the motors always remain bound to the cargo. When opposite-polarity motors engage simultaneously 
with the track, each filament-bound motor exerts force on the cargo in their respective direction, resulting in a net force on the cargo in one of the 
directions. This net force experienced by both sets of motors is called  `load' and is generally assumed to be shared equally among all the motors 
which move in same direction. It is now well-established that the detachment rates $\epsilon_{\pm}$ depend on the load per motor, while the attachment 
rates are generally found to be independent of load. Based on Kramers rate theory, it is generally assumed that this load-dependence of 
detachment rates is exponential, i.e., $\epsilon_{\pm}(f)=\epsilon_{\pm}(0)\exp(f/f^{d}_{\pm})$, where $f_{\pm}^{d}$ is usually called the detachment 
force and $f$ is the load per motor. However, recent investigations\cite{Deepak,Kunwar,phagosome} have shown that the load-dependence of the 
dissociation rate of dynein (but not kinesin\cite{phagosome}) deviates significantly from exponential behavior in the super-stall regime. Therefore, 
we adopt the exponential load-dependence for kinesin's detachment rate, whereas for dynein, we assume exponential dependence up to the stall force, 
beyond which the rate is insensitive to load\cite{Deepak}. This model is roughly consistent with in vitro experimental observations\cite{Kunwar}, and 
has been attributed to a catch-bond situation in the motor-filament interaction\cite{Kunwar}. Further details of the model, especially regarding its 
implementation in numerical simulations may be found in \cite{Muller}, as well as our earlier paper\cite{Deepak}.
 
The load-dependence of the detachment rates significantly affects the properties of bidirectional cargo motion, and is a necessary feature in the 
model so as to reproduce experimentally observed features of the saltatory motion of cargoes, e.g., lipid droplets in {\it Drosophila}\cite{Muller}. 
Nevertheless, it turns out from our study that it is not crucial to understand the origin of memory in bidirectional transport. For this reason, we 
first develop our formalism with load-independent detachment rates and include load dependence in numerical simulations in the later stages, where 
we study biologically relevant situations. As it turns out, load-dependence of motor detachment rate introduces only a quantitative modification of the 
parameters of interest in this context. 

In the stochastic TOW model, with the elapse of time, the number $d(k)$ of actively hauling dyneins (kinesins) change, such that $0\leq d\leq D$ and $0\leq k \leq K$. 
It may be noted that for a given source state $(d,k)$ of the cargo, there are between two to four possible target states: $(d,k\pm 1)$ and $(d\pm 1,k)$, 
subject to the bounds above. Consequently, the cargo switches between different states: plus-moving state(s) when only kinesins are active, 
minus-moving state(s) when only  dyneins are active, TOW state(s) when both kind of motors are active together and finally, the detached state when 
all the motors are inactive on the cargo\cite{Klumpp,SKlumpp,Muller,Mjmuller}. In fig.\ref{fig:fig1}(b) corresponding motility states of cargo are 
shown for two simple cases $(D=1,K=1)$ and $(D=2,K=1)$. We should notice that, when two dyneins and a kinesin are hauling a cargo, both minus-run 
($m_1$ and $m_2$) and TOW state ($\tau_1$ and $\tau_2$) become {\it degenerate}. In general, both run and pause states of the cargo become degenerate 
when more than one motor is used in one or either directions.

Let $m=\{m_i\}$ represent the set of all minus-moving states of the cargo, $p=\{p_i\}$ represent all plus-moving states, $\tau=\{\tau_i\}$ represent 
all TOW states and $\{o\}$ represent the completely detached state. Let $\phi(m,\tau|p)$ be the probability that a cargo in plus-run enters a TOW 
(without detaching from the filament) and then switches direction, with unspecified durations spent in plus run and TOW; this may hence be defined as 
the direction reversal probability for the plus state, while $\phi(p,\tau|m)$ gives the same for the minus state. Then, direction preserving 
probabilities are given by the normalization conditions 
\begin{equation}
\phi(p,\tau|p)+\phi(m,\tau|p)=1=\phi(m,\tau|m)+\phi(p,\tau|m). 
\label{eq:eq1}
\end{equation}

Given these probabilities, we define the memory parameter\cite{Goldstein} 
\begin{equation}
\mu\equiv \phi (p,\tau|p)-\phi (p,\tau|m)=\phi (m,\tau|m)-\phi(m,\tau|p),
\label{eq:memory}
\end{equation}

such that $\mu=0$ means no memory, while $\mu \neq 0$ means that the probability of finding the cargo in a certain run state is dependent on its run 
direction preceding the TOW. Memory of direction, as defined above, is distinct from a possible overall bias in the motion of cargo towards plus or 
minus directions. The bias parameter may be defined as

\begin{equation}
\nu\equiv \gamma_{p}-\gamma_m,
\label{eq:eq1++}
\end{equation}

where $\gamma_{p}=\Sigma_{i} \gamma_{p_i}$ and $\gamma_{m}=\Sigma_{i} \gamma_{m_i}$ give the probabilities for a TOW to terminate in plus and minus run, 
respectively. Intuitively, it appears likely that memory must affect the bias in transport, however no general relation between the two is known, to 
the best of our knowledge. Nevertheless, it can be shown that, if the probability of cargo dissociating completely from microtubule is very small, 
$\nu\simeq \eta/(1-\mu)$ (see Eq.\ref{eq:eq12} later, and also Appendix B), where we define 

\begin{equation}
\eta\equiv \phi(p,\tau|p)-\phi(m,\tau|m),
\label{eq:eq1+}
\end{equation}

 as the asymmetry coefficient in the transport. In the present problem, however, there is a non-zero probability for the cargo to completely detach 
from the filament, and therefore the above relation holds only in situations where this can be neglected. 

If the plus and minus run lengths  are constants and have the same value, then $\nu$ alone determines the average direction of motion of the cargo, 
i.e., the sign of the drift velocity $v_d\equiv \lim_{t\to\infty}t^{-1}\langle x\rangle$. This is a standard assumption in most mathematical studies 
of the persistent random walk, but is unrealistic in our case, as the plus and minus run durations are strongly dependent on the binding and unbinding 
rates of the motors\cite{Deepak}, which also determine $\mu$ and $\nu$. In the present paper, we focus on memory, bias and directional correlations as 
appropriate to a random walk-like picture of the motion; a complete characterization of bidirectional motion also requires identification of the 
different regimes of transport as well as a detailed study of drift and diffusion coefficients of the walk. We certainly hope to address these issues 
in a later publication. 

We now develop the mathematical formalism required to derive explicit expressions for the memory parameter. From a state $\alpha$ of a cargo, the 
probability of transition to another state $\beta$ at a time between $t$ and $t+dt$ is $F_{\alpha\beta}(t) dt =r_{\alpha\beta}\psi_{\alpha}(t) dt$, 
where $r_{\alpha\beta}$ is the (constant) rate for the $\alpha\to \beta$ transition and $\psi_\alpha(t)=e^{-t\sum_{\beta}r_{\alpha\beta}}$ is the {\it 
survival probability}, defined as the probability for the cargo to stay in state $\alpha$ during the interval $[0:t]$. Because the active/inactive 
configuration of cargo-bound motors determine the state $\alpha$, $\psi_\alpha(t)$ can be expressed in terms of survival probability of a motor in 
active state, i.e., $e^{-\epsilon_{\pm}t}$ and inactive state, i.e.,  $e^{-\pi_{\pm}t}$, on the cargo. The probability of the transition is 
$\Phi (\beta|\alpha;t)=\int^{t}_{0}F_{\alpha\beta}(\tau) d\tau$, and its steady state limit $\Phi(\beta|\alpha;t\to\infty)\equiv \Phi(\beta|\alpha)$, 
is a two-point Green function (`bare' propagator), which is fundamental to our analysis. One may, similarly, define a 3-point Green function 
$\Phi (\gamma,\beta|\alpha;t)$ in the problem, i.e., the probability for the system to trace a certain a history of states $(\alpha,\beta,\gamma)$ 
during a time interval $[0:t]$, having started from $\alpha$ at $t=0$. Given the Markovian nature of the underlying process, this probability is 
expressed in the form of a convolution in time: $\Phi (\gamma,\beta|\alpha;t)=\int_{0}^{t}d\tau_1F_{\alpha \beta}(\tau_1)\int_{0}^{t-\tau_1}d\tau_2F_{\beta \gamma}(\tau_2)$. 
It follows that, in the long time limit, the steady state probability $\Phi (\gamma,\beta|\alpha;t\to\infty)\equiv \Phi(\gamma,\beta|\alpha)$ is 
expressed as the product:
\begin{equation}
\Phi (\gamma,\beta|\alpha)=\Phi(\gamma|\beta) \Phi(\beta|\alpha).
\label{eq:eq3}
\end{equation}

It is now important to define a set of generalized two-point Green functions $G(\beta|\alpha)$, which, for ease of distinction, we shall refer to as 
`dressed' propagators. The difference between the bare and dressed propagators may be explained with an example: whereas $\Phi(\tau_j|m_i)$ is  the 
probability for the cargo to be in a TOW state $\tau_j$, after having spent an unspecified duration of time in the minus-moving state $m_i$  (with no 
other state transitions in between), $G(\tau_j|m_i)$ includes an indefinite number of cyclic transitions between the various (degenerate) $\{m_i\}$ 
states, but fixed initial and final states $m_i$ and $\tau_j$. From both types of propagators, higher order Green functions may be constructed using 
Eq.\ref{eq:eq3}; see examples below.

\section{Memory in Cargo transport}
\subsection{Exact results}
{\it  Case(i) $D=1 ~K=1$:} Fig. \ref{fig:fig1}(b) (panel 1) shows list of four possible states of cargo ($o, p_1,m_1$ and $\tau_1$), when it is driven 
by a kinesin and a dynein. The corresponding survival probabilities are $\psi_{m_1}(t)=e^{-(\epsilon_-+\pi_+)t},\psi_{p_1}(t)=e^{-(\epsilon_+ + \pi_-)t}, \psi_{\tau_1}(t)=e^{-t\Sigma_\epsilon}$ 
and $\psi_o(t)=e^{-t\Sigma_{\pi}}$  respectively, where, for later convenience, we have introduced the compact notations $\Sigma_{\epsilon}\equiv \epsilon_{+}+\epsilon_{-}$ 
and $\Sigma_{\pi}\equiv \pi_{+}+\pi_{-}$. The two-point Green functions immediately follow:
\begin{eqnarray}
\Phi(\tau_1|m_1)=\frac{\pi_+}{\pi_++\epsilon_{-}}~~;~~\Phi(\tau_1|p_1)=\frac{\pi_-}{\pi_-+\epsilon_{+}} ; \nonumber\\
\Phi(m_1|\tau_1)=\frac{\epsilon_+}{\Sigma_\epsilon}=1-\Phi(p_1|\tau_1).~~~~~~~~~~~~~~~~~~
\label{eq:eq3+}
\end{eqnarray}
 The direction reversal of plus-run through a TOW  corresponds to the transition path $p_1 \rightarrow \tau_1 \rightarrow m_1$ while that of minus-run 
corresponds to the path $m_1 \rightarrow \tau_1 \rightarrow p_1$, with $\Phi (m_1,\tau_1|p_1)$ and $\Phi (p_1,\tau_1|m_1)$ being the respective 
three-point Green functions representing the processes. On the other hand, $p_1 \rightarrow \tau_1 \rightarrow p_1$ and $m_1 \rightarrow \tau_1 \rightarrow m_1$ 
paths correspond to direction preserving  transitions through a TOW of a plus and minus run respectively (with Green functions $\Phi (p_1,\tau_1|p_1)$ 
and $\Phi (m_1,\tau_1|m_1)$). It is convenient to normalize the three point functions as below:
\begin{equation}
\phi(m,\tau|p)=\frac{\Phi(m_1,\tau_1|p_1)}{\sum_{\alpha}\Phi(\alpha,\tau_1|p_1)};
\phi(p,\tau|m)=\frac{\Phi(p_1,\tau_1|m_1)}{\sum_{\alpha}\Phi(\alpha,\tau_1|m_1)},
\label{eq:eq4}
\end{equation}
where $\alpha=\{p_1,m_1\}$. After carrying out the required calculations using Eq. \ref{eq:eq3}-\ref{eq:eq4} we find that  
\begin{eqnarray}
\phi (p,\tau|p)=\frac{\epsilon_-}{\Sigma_\epsilon}=\phi (p,\tau|m),
 \label{eq:eq5}
\end{eqnarray}
and by normalization (Eq. \ref{eq:eq1}) we can write $\phi (m,\tau|m)=\phi(m,\tau|p)$. Therefore, by definition cargo motion is memoryless ($\mu =0$) 
and hence, random walk exhibited by a cargo driven by a dynein and a kinesin is Markovian. Note, however, that if $\epsilon_- < \epsilon_+$, 
$\eta<0$ so cargo is biased towards minus direction, on the other hand if $\epsilon_- > \epsilon_+$, then $\eta>0$ so cargo is biased towards plus 
direction locally. 

{\it Case(ii) $D=2 ~K=1$:} As shown in Fig. 1(b) (panel 2), cargo driven by two dyneins and a kinesin has two minus-moving states ($m_1$ and $m_2$), 
two TOW states ($\tau_1$ and $\tau_2$), a single plus-moving state ($p_1$) and a detached state ($o$). The corresponding survival probabilities are 
$\psi_{m_1}(t)=e^{-(\epsilon_- + \Sigma_\pi)t}, \psi_{m_2}(t)=e^{-(2\epsilon_- + \pi_+)t}, \psi_{\tau_1}(t)=e^{-(\pi_- + \Sigma_\epsilon)t}, \psi_{\tau_2}(t)=e^{-(\epsilon_- + \Sigma_\epsilon)t}, \psi_{p_1}(t)=e^{-(\epsilon_+ + 2\pi_-)t}$ 
and $\psi_o(t)=e^{-(\pi_- + \Sigma_\pi)t}$. From these probabilities, and with the identification of the relevant rates, the bare propagators follow:
\begin{eqnarray}
\Phi(m_2|m_1)=\frac{\pi_{-}}{\epsilon_{-}+\Sigma_\pi}~;~\Phi(\tau_1|m_1)=\frac{\pi_{+}}{\epsilon_{-}+\Sigma_\pi};\nonumber\\
\Phi(m_1|m_2)=\frac{2\epsilon_{-}}{2\epsilon_{-}+\pi_+}=1-\Phi(\tau_2|m_2);\nonumber\\
\Phi(\tau_2|\tau_1)=\frac{\pi_{-}}{\pi_{-}+\Sigma_\epsilon}~;~\Phi(m_1|\tau_1)=\frac{\epsilon_{+}}{\pi_{-}+\Sigma_\epsilon};\nonumber\\
\Phi(p_1|\tau_1)=\frac{\epsilon_{-}}{\pi_{-}+\Sigma_\epsilon}~;~\Phi(\tau_1|p_1)=\frac{2\pi_{-}}{\epsilon_{+}+2\pi_{-}}; \nonumber\\
\Phi(m_2|\tau_2)= \frac{\epsilon_+}{\epsilon_-+ \Sigma_\epsilon}=1-\Phi(\tau_1|\tau_2).
\label{eq:eq5+}
\end{eqnarray}
However, because of the possibility of cyclic transitions between degenerate states, in the present case, the dressed propagators $G(\beta|\alpha)$ 
are more useful, which are related to their bare counterparts through equations analogous to Dyson's equation in quantum field theory. For example, 
it is easily seen that, for $j=1,2$, $G(\tau_j|m_j)=\Phi(\tau_j|m_j)(1+\Omega_m+\Omega^2_m+.....)$, where $\Omega_m=\Phi(m_1|m_2)\Phi(m_2|m_1)$ is the 
probability for the cyclic transition $m_1\to m_2\to m_1$. We may similarly define $\Omega_{\tau}=\Phi(\tau_1|\tau_2)\Phi(\tau_2|\tau_1)$ as the 
probability for the $\tau_1\to \tau_2\to \tau_1$ cyclic transition. The complete set of such relations are given below:
\begin{eqnarray}
G(\tau_j|m_j)=\frac{\Phi(\tau_j|m_j)}{1-\Omega_m}~;~G(\tau_i|m_j)=\frac{\Phi(\tau_i|m_i)\Phi(m_i|m_j)}{1-\Omega_m};~~~\nonumber\\
G(m_j|\tau_j)=\frac{\Phi(m_j|\tau_j)}{1-\Omega_{\tau}}~~;~~ G(m_i|\tau_j)=\frac{\Phi(m_i|\tau_i)\Phi(\tau_i|\tau_j)}{1-\Omega_{\tau}};~~~\nonumber\\
G(p_1|\tau_1)=\frac{\Phi(p_1|\tau_1)}{1-\Omega_{\tau}}~~;~~G(p_1|\tau_2)=\frac{\Phi(p_1|\tau_1)\Phi(\tau_1|\tau_2)}{1-\Omega_{\tau}},~~~~~
\label{eq:3++}
\end{eqnarray}
where $i,j=1,2$ and $i\neq j$. However, $G(\tau_1|p_1)=\Phi(\tau_1|p_1)$ while $G(\tau_2|p_1)=0$ (A slightly different, and alternative method for 
computing the propagators is described in Appendix A).

The complete set of (dressed) three-point Green functions are then constructed from these propagators as $G(\gamma,\beta|\alpha)=G(\gamma|\beta)G(\beta|\alpha)$, 
following Eq.\ref{eq:eq3}. It is then easily seen that the direction reversal probabilities for the plus and minus states are given by
\begin{eqnarray}
\phi(m,\tau|p)=\frac{\sum_{\ell} G(m_\ell,\tau_1|p_1)}{\sum_{\ell} G(m_\ell,\tau_1|p_1)+G(p_1,\tau_1|p_1)},~~~~~~~~~~\label{eq:eq6a}\\
\phi(p,\tau|m)=\frac{\sum_{i,j}\gamma_{m_i} G(p_1,\tau_j|m_i)} {\sum_{i}\gamma_{m_i}\bigg[\sum_jG(p_1,\tau_j|m_i)+\sum_{j,\ell}G(m_\ell,\tau_j|m_i)\bigg]},~~~~
\label{eq:eq6b}
\end{eqnarray}
where all the indices $i,j, \ell=1,2$ and $\gamma_{m_i}$ is the probability to find the system in state $m_i$ after TOW. It is convenient to 
define the ratios $\eta_i=\gamma_{m_i}/\gamma_{p_1}$, where $\gamma_{p_1}=1-\sum_{i=1,2}\gamma_{m_i}$ is the probability to be in plus-moving state 
after a TOW. The coefficients $\eta_i$ are now determined using the self-consistency conditions

\begin{equation}
\sum_{i,j} \eta_iG(m_\ell,\tau_j|m_i)+G(m_\ell,\tau_1|p_1)=\eta_\ell~~;~~\ell=1,2.
\label{eq:eq7}
\end{equation}

Eq. \ref{eq:eq6a} and Eq.\ref{eq:eq6b} reduce to Eq.\ref{eq:eq4} when there is no degeneracy in $m$ and $\tau$ states. The direction preserving 
probabilities are then found from normalization (Eq. \ref{eq:eq1}). Using Eq. \ref{eq:eq5+}-\ref{eq:eq7}, we obtain the following explicit expressions 
for $\phi (p,\tau|p)$ and $\phi (p,\tau| m)$, analogous to Eq. \ref{eq:eq5}:

\begin{widetext}
\begin{eqnarray}
\phi (p,\tau|p)=\frac{\epsilon_- ( \epsilon_+ + 2 \epsilon_-)}{\epsilon^2_ + + (3\epsilon_- + \pi_-)\epsilon_+  + 2 \epsilon^2_- }, \nonumber\\
\phi (p,\tau| m)=\frac{2 \epsilon^2_- \Big[   (\epsilon_+  + 2 \epsilon_- + \pi_-)(\epsilon_+ + \pi_+ + 2 \epsilon_- + \pi_-)  + \epsilon_- \pi_-    \Big ]}
{\Big[(2 \epsilon_- + \pi_-)(\epsilon_+ + \pi_+ + 2 \epsilon_- +\pi_-) + \epsilon_- \pi_- \Big ] \Big [\epsilon^2_ + + (3\epsilon_- + \pi_-)\epsilon_+  + 2 \epsilon^2_-   \Big ]}.
\label{eq:eq11}
\end{eqnarray}
\end{widetext}

\begin{figure}
\includegraphics[width=8.5cm,keepaspectratio=true]{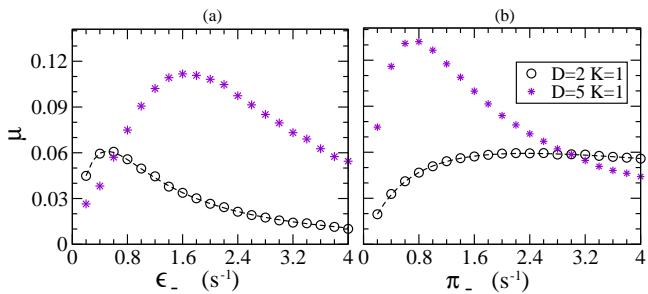}
\caption{\label{fig:fig2} (color online) Non-zero value of $\mu$ when it is plotted as a function of (a) the unbinding rate $\epsilon_-$  and (b) the binding rate 
$\pi_-$ of dynein shows the presence of memory in the cargo transport.  The dashed line represents the theoretical curve for $(D=2,K=1)$ case 
(Eq. \ref{eq:eq11}) while the symbols are numerical results for $(D=2,K=1)$ and $(D=5,K=1)$ cases. The parameters $\pi_+=0.904 s^{-1}, \epsilon_+=0.314 s^{-1}$ 
and $\pi_-=2.74 s^{-1}$ in (a) and $\epsilon_-=0.667 s^{-1}$ in (b) are fixed using in vitro experimental data given in Table\ref{tab:tab1}.}
\end{figure}
 
Clearly, $\mu$ is non-zero in this case, indicating that the random walk is non-Markovian in nature. Here, although the survival probability in each 
individual motor state is an exponentially decaying function, the survival probability of the cargo in minus-run or TOW  is modified by internal 
transitions between degenerate states ($m_1$ and $m_2$  or  $\tau_1$ and $\tau_2$). By constructing the master equation for these degenerate states, 
one can show that the waiting time distribution in minus-run/TOW is a multi-exponential function\cite{supp}. This is similar to the recent discovery 
of non-Markovian behavior in enzyme kinetics characterized by multi-exponential waiting time distributions, when more than one enzyme is present in 
the system \cite{Ronojoy}. Because memory originates from degeneracy of states, it is natural to expect that $\mu$ will be non-zero in the more general 
$K>1, D>1$ cases also. 

\subsection{Numerical simulations}
To support our mathematical results, to explore higher values of $D$ and $K$ and to consider the effects of load-dependence of detachment rates, we 
next performed numerical simulations using a Gillespie algorithm \cite{Gillespie} for several cases of multiple motor transport $(D\leq 20, K\leq 2)$. 
For details of the simulations including the fixing of binding and unbinding rates and determination of the velocity of cargo motion, the reader is 
referred to our earlier paper\cite{Deepak}.

{\it Memory with load-independent detachment rates:} For  $(D=2,K=1)$ and  $(D=5,K=1)$ cases, $\mu$ is plotted as a function of dynein unbinding rate 
$\epsilon_-$ and binding rate $\pi_-$ in Fig. \ref{fig:fig2}(a) and (b) respectively. For large and very small values of  $\epsilon_-$ or  $\pi_-$, 
the cargo stays mostly in one of the extreme degenerate states thereby reducing the effect of degeneracy and hence, the memory is smaller. For 
intermediate values, on the other hand, the transitions between the degenerate states of minus-run or TOW are much more frequent, and this leads to 
maximization of memory.

\begin{figure}
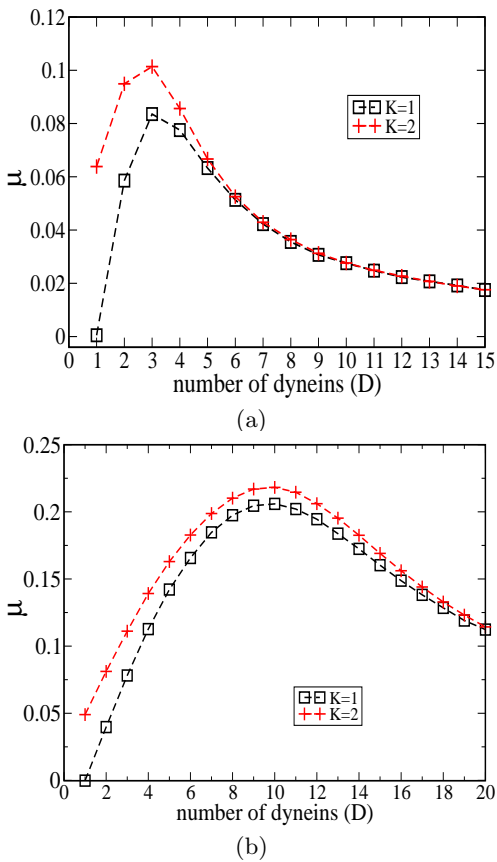

\includegraphics[width=6.5cm,height=5.2cm]{fig3a.eps}\\
(a)\\
\includegraphics[width=6.5cm,height=5.2cm]{fig3b.eps}\\
(b)\\
\caption{\label{fig:fig3} (color online) Memory parameter ($\mu$) as a function of number of dyneins is plotted in (a) with load independent and in (b) with load 
dependent detachment rates for motors. In both cases, the memory appears to vanish for very large and very small number of bound dyneins. But $\mu$ is 
two-fold larger when detachment rates are load dependent (details in text).}
\end{figure}

To investigate the memory effect more extensively, we studied the memory parameter as a function of the motor numbers (see Fig.\ref{fig:fig3}(a)), 
keeping the binding/unbinding rates of dynein and kinesin at fixed values which are given in Table\ref{tab:tab1}. An increase in the number of 
dyneins increases the number of degenerate  states and hence $\mu$ increases initially. However, for large number of bound dyneins, the central limit 
theorem comes into play, and the cargo is now found with overwhelmingly large probability in one of the degenerate states, corresponding to the 
average number of bound motors. Therefore, the effective number of degenerate states is now smaller, leading to a reduction in $\mu$. 

\begin{table}
\begin{tabular}{|c|c|c|c|}
\hline 
Molecular motor &$\epsilon_{\pm}~~~(s^{-1})$&$\pi_{\pm} ~~~ (s^{-1})$&$f^{d}_{\pm}~~~ (pN)$ \\\hline
Kinesin(+)      &0.314                      &0.904                   &5.169\\\hline
Dynein($-$)     &0.667                      &2.740                   &0.546\\\hline
\end{tabular}
\caption{List of single-molecule parameters extracted from a previous in vitro study\cite{Soppina}. A detailed discussion is to be found in \cite{Deepak}. 
However, other studies have reported different binding and unbinding rates for dynein and kinesin \cite{Leidel,phagosome,KSvoboda,RDVale,Coppin,MAWelte,MJSchnitzer,SToba,Vaishnavi}.}
\label{tab:tab1}
\end{table}

{\it Memory with load-dependent detachment rates:} We will now address the question of how the load-dependence of detachment rates of the motors 
affect the memory parameter. Simulations show that, here, the dependence of $\mu$  on the number of dyneins (see Fig.\ref{fig:fig3}(b)) is qualitatively 
the same as in the load-independent case studied in Sec.III B. However, for the present choice of parameters, the memory parameter is almost two-fold 
larger while the maximum is shifted to larger dynein numbers. 

\begin{figure}
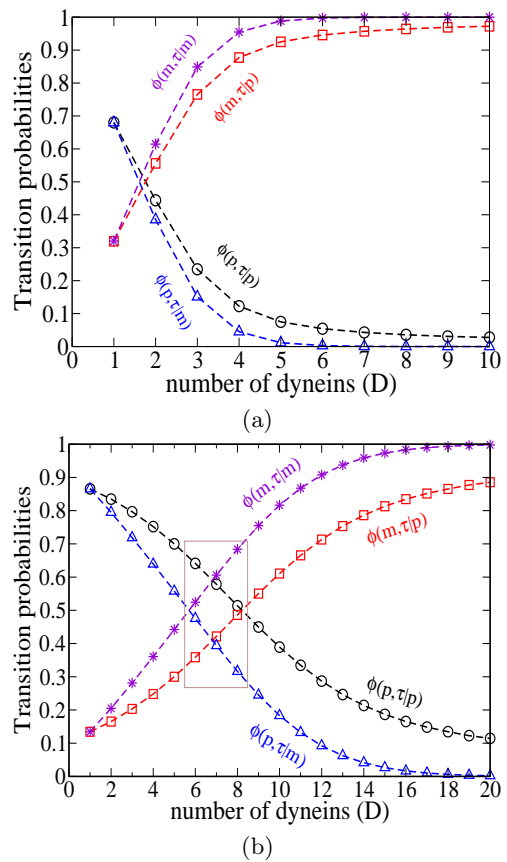

\includegraphics[width=6.5cm,height=5.2cm]{fig4a.eps}\\
(a)\\
\includegraphics[width=6.5cm,height=5.2cm]{fig4b.eps}\\
(b)\\
\caption{\label{fig:fig4} (color online) Direction reversal and preserving probabilities of cargo as a function of number of dyneins, with $K=1$ is shown here, 
when detachment rates of motors are assumed to be load independent (a) and load dependent (b).  In (b), inside the box, for $D=6, 7$ and $8$, general 
persistence of direction is observed, i.e., both plus and minus-directed cargoes are more likely to continue moving in the same direction after a TOW.}
\end{figure}

In Fig.\ref{fig:fig4}(a) and  Fig.\ref{fig:fig4}(b), we have plotted the direction reversal and direction preserving probabilities individually, as 
a function of the number of dyneins $D$, and fixing $K=1$. For $D=6, 7$ and $8$ (Fig.\ref{fig:fig4}(b), inside the box), both plus and minus-directed 
cargoes are more likely to continue moving in the same direction after a TOW {\it i.e.} $\phi(p,\tau|p)> \phi (m,\tau|p)$ and $\phi (m,\tau|m) > \phi (p,\tau|m)$, 
therefore the bidirectional motion becomes a persistent random walk. Persistence in cargo motion was observed  in the experiments of Leidel et al.\cite{Leidel} 
suggesting the presence of memory in transport; however the reasons are likely to be different because of the presence of the trap. 

{\it Bias in cargo transport:} 
Local preference towards one of the directions after a TOW is quantified by the difference between the probability that the TOW terminates in plus 
and minus directions and is characterized by the bias parameter $\nu$, defined in Sec.II. An approximate relation can be shown to exist between the 
bias $\nu$, memory parameter $\mu$ and the coefficient of asymmetry $\eta$, under conditions where the probability of complete detachment of the 
cargo from the filament is small. This is derived in Appendix B (for the special case $D=2,K=1$), and the final result is

\begin{equation}
\nu\simeq \frac{\eta}{1-\mu},
\label{eq:eq12}
\end{equation}

Two features are noteworthy in the above expression:(i) bias has the same sign as the asymmetry coefficient and (ii) for fixed $\eta$, bias is 
enhanced by positive memory of direction ($\mu>0$) and suppressed by negative memory ($\mu<0$). 

\begin{figure}
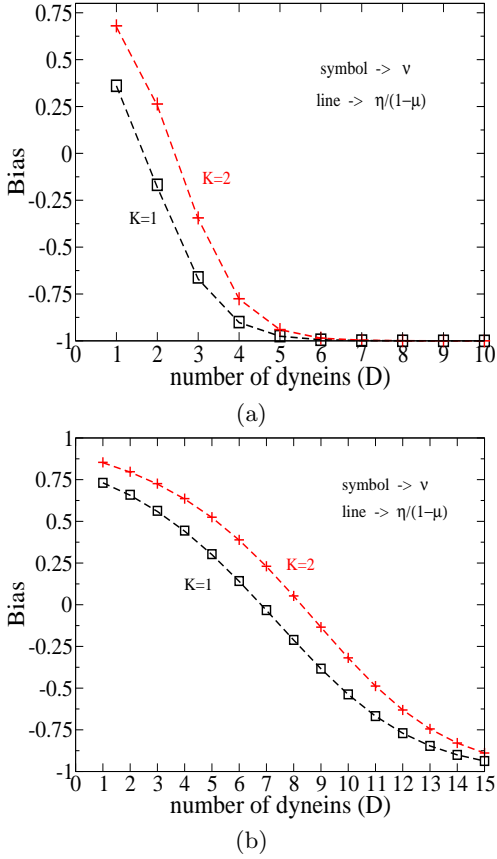

\includegraphics[width=6.5cm,height=5.2cm]{fig5a.eps}\\
(a)\\
\includegraphics[width=6.5cm,height=5.2cm]{fig5b.eps}\\
(b)\\
\caption{\label{fig:fig6} (color online) Symbols (black square and red plus) show the directly measured bias parameter $\nu$ (Eq.\ref{eq:eq1++}) as a function of 
number of dyneins (a) without and (b) with load dependent detachment rates for motors, as measured in simulations. The lines give the predicted values 
using Eq.\ref{eq:eq12}, with $\eta$ and  $\mu$ measured separately in simulations.}
\end{figure}

Fig. \ref{fig:fig6} shows $\nu$, as defined in Eq.\ref{eq:eq1++}, computed using the probabilities $\gamma_{p}$ and $\gamma_m$ measured in simulations, 
plotted as a function of the number of dyneins, when detachment rates of motors are assumed to be load-independent (a) and load-dependent (b). Not 
surprisingly, an increase in the number of attached dyneins, with fixed number of kinesins, leads to eventual reversal in the sign of the bias 
parameter from plus to minus; with load-dependence of detachment rates, this reversal occurs at higher dynein numbers. A comparison with the 
approximate expression in Eq.\ref{eq:eq12} is also shown in each figure. Here, the values of $\eta$ and $\mu$ are computed numerically using the 
direction reversal probabilities determined from simulations. In both cases, the approximate relation in Eq.\ref{eq:eq12} manages to capture the 
observed variation very well, and it therefore appears that it is of general validity, beyond the specific motor number combination for which it was derived. 
 
In the present context,  is also important to note that a non-zero bias is not necessary for net drift of the cargo in one direction because the run 
durations can be different in each direction. For example (under no-detachment conditions for the cargo), in the $(D=1,K=1)$ case, we find that 
$\nu=(\epsilon_- -\epsilon_+)/(\epsilon_- +\epsilon_+)$ exactly, which is independent of the binding rates $\pi_{\pm}$. Therefore, when 
$\epsilon_+=\epsilon_-$, the cargo shows unbiased motion ($\nu=0$)  whereas, from symmetry reasons, it will clearly have non-zero average velocity 
(drift) if $\pi_+ \neq \pi_-$. In this case, although plus and minus directions are equally favored after a TOW, the time durations spent by the cargo 
in plus or minus run depend on $\pi_{\pm}$\cite{DAE},  which in turn leads to non-zero drift. 
 
\subsection{Correlation in run directions}

A direct measure of correlations between the directions of runs, separated by one or more TOW events, is the directional correlation function $C(n)$, 
which we define as follows:

\begin{equation}
 C(n)= \lim_{i\to\infty}\frac{\langle S_i S_{i+n}\rangle-\langle S_i\rangle\langle S_{i+n}\rangle}{\langle S_i^2\rangle-\langle S_i\rangle^2} ~;~n\geq0,~~~~~~
\label{eq:eq13}
\end{equation}

where $S_i=1$ if the cargo runs in the plus direction after the $i$'th TOW event, and $S_i=-1$ if it runs in the minus direction. The second term in 
the numerator takes out the effect of the bias, and the denominator is a normalization factor, introduced such that $C(0)=1$. The directional correlation 
function is analogous to the standard velocity autocorrelation function, but with certain important differences. A detailed treatment of the latter 
is not the subject of this paper, but a brief discussion is given in the supplementary material\cite{supp}.

We now make a conjecture that the directional correlation function decays exponentially with the number of TOWs, i.e., $C(n)=\rho^n$ where 
$0<\rho<1$\cite{remark}. In Appendix C, we have shown for $(D=2, K=1)$ case that, under conditions where complete detachment of the cargo from the 
filament is neglected, $C(1)=\mu$ exactly. However, this can be generalized to other cases i.e., {\it $(D\geq 2, K\geq 2)$} also. Therefore, under 
this approximation, we arrive at the simple and interesting result that $\rho=\mu$, and hence

\begin{equation}
C(n)\simeq \mu^n=e^{-\frac{n}{n_c}}~~~n\geq 0.
\label{eq:eq14}
\end{equation}

where $n_c=-(\ln \mu)^{-1}$ is the equivalent of a correlation time, and gives the mean number of TOWs over which directional correlation between runs 
persists. This is indeed an intuitively pleasing relation as this clearly shows that the number of TOWs over which the directional correlations are 
appreciable increases with $\mu$ (For $\mu=0$, $C(n)=0$ for all $n\geq 1$). 

Table \ref{tab:tab2} gives a comparison between $C(n)$ directly measured in simulations versus the prediction in Eq.\ref{eq:eq14} for $1\leq n\leq 3$, 
and varying dynein numbers $D$ (keeping $K=1$). It is clear that Eq.\ref{eq:eq14} approximates the simulation data rather well. 

\begin{table}
\begin{tabular}{|c|c|c|c|c|}
\hline 
motor configuration &$\mu$      &$n$&   $C(n)$ (sim) &Eq.\ref{eq:eq14} \\\hline
                    &           &1&     0.00041&      0 \\ 
(D=1,K=1)           &0          &2&     0.00003&      0\\
                    &           &3&    -0.00004&      0\\\hline
                    &           &1&     0.14192&      0.14213\\ 
(D=5,K=1)           &0.14213    &2&     0.02196&      0.02020\\
                    &           &3&     0.00350&      0.00287\\\hline
                    &           &1&     0.20359&      0.20592\\ 
(D=10,K=1)          &0.20592    &2&     0.04333&      0.04240\\
                    &           &3&     0.00925&      0.00873\\\hline
                    &           &1&     0.15860&      0.16026\\ 
(D=15,K=1)          &0.16026    &2&     0.02639&      0.02568\\
                    &           &3&     0.00434&      0.00411\\\hline
                    &           &1&     0.11047&      0.11234\\ 
(D=20,K=1)          &0.11234    &2&     0.01342&      0.01262\\
                    &           &3&     0.00129&      0.00141\\\hline
\end{tabular}
\caption{The table shows the directional correlation function, defined in Eq.\ref{eq:eq13}, as measured in simulations, compared  to the prediction 
of Eq.\ref{eq:eq14}, when detachment rates of motors are load dependent. The $\mu$ values were separately found in simulations (data plotted in 
Fig.\ref{fig:fig3}b). The reported values of $C(n)$ were obtained by averaging over $5\times 10^7$ independent cargo trajectories.}
\label{tab:tab2}
\end{table}

\section{Conclusions and Discussion}

Random walk models have found a large number of applications in modeling various kinds of biological transport phenomena (see e.g., \cite{codling} 
for a review). The bidirectional transport of cargoes like mitochondria, lipid droplets, endosomes, phagosomes etc in eukaryotic cells is also akin 
to a random walk on microtubule filaments. It is clear that for functional reasons, this walk is likely to be biased; certain cargoes need to be moved 
to the interior of the cell, while certain others may need to be transported to the outer cell membrane. The underlying molecular mechanisms of 
bidirectional transport have been studied in a great deal of detail, both experimentally and via biophysical modeling. However, barring a few papers, 
much less attention has been paid to the statistical properties of the walk itself, which motivated us to undertake this study.

Our focus here was on understanding correlations between successive run directions of a cargo moving bidirectionally, being transported by two 
opposing teams of kinesins and dyneins. By studying a memory parameter constructed using direction reversal probabilities, we showed that memory in 
direction is a generic property of this motion, which appears when at least one team of motors has more than one member. TOW between two single 
motors on either side, however, results in a biased random walk of the cargo without memory. Interestingly, the memory parameter is  found to be a 
non-monotonic function of the number of motors and, for fixed binding/unbinding rates, is maximized for a certain motor number. We also find that the 
effective interaction between the opposing motor teams which emerges out of the load-dependence of the individual unbinding rates, enhances this 
memory. For one set of experimentally measured binding and unbinding rates for dynein and kinesin (in vitro studies using motor proteins from {\it D. 
discoideum}, see Table \ref{tab:tab1}), we estimated that the correlation in run direction could persist up to 2-3 TOW events for typical motor 
numbers (in this case, the upper limit corresponds to 1-2 kinesins and 8-12 dyneins). 

Correlated random walks with memory and persistence have been the subject of a large number of mathematical studies\cite{furth,taylor,Goldstein,patlak,weiss,Ricardo,hermann}. 
In the models studied in these papers, at each instant, the walker takes a step of fixed size in a certain direction, the probability for which 
depends on one (usually) or more previous steps that have been taken. It has been shown that, in one dimension, after appropriate limiting procedures, 
the equation that describes the asymptotic properties of such a walk is the telegraphers equation\cite{Goldstein,weiss}, which reduces to (a) the 
standard diffusion equation in the long-time limit, and (b) the wave equation in the short-time limit. The demarcation of these regimes is determined 
by the time scale over which the steps remain correlated. The bidirectional transport model studied in this paper, clearly falls under the class of a 
correlated random walk, but with some additional and unique features: (a) the run duration, equivalent to the step size of the random walk, is not a 
constant, but determined by binding and unbinding rates of the motors (b) the TOW/pause state can have a non-zero velocity (but small compared to 
run states) depending on the number of opposing motors (c) the cargo may detach as a whole from the filament, which rules out steady state behavior of 
the Green functions, even in the long-time limit. It would be interesting to see if a continuum equation, analogous to the telegraphers equation, 
could be constructed for the present problem in the long-time limit, taking into account these modifications, and to study its properties. It is also 
pertinent to note that given the finite size of the cell, the biologically relevant time regime need not necessarily be the long-time limit mentioned 
above, but could be the memory-dominated short-time limit. If this is true, the presence of multiple motors in a team could function as a mechanism to 
provide a semi-deterministic character to the cargo motion. The implications of this conjecture, as well as its systematic verification remain to be 
done and are among our future goals. 

The mathematical and computational results in this paper should be verifiable in experiments. Detailed time-traces of cargo trajectories in 
bidirectional motion have been obtained from both in vitro and in vivo experiments\cite{Leidel,Kunwar,Soppina,Hendricks}. By analyzing such trajectories, it 
should be possible to measure the memory parameter $\mu$ and correlate it with the number of motors estimated by other means (eg. optical trap stalls). 
In in vivo situations, our results and methods may be found useful in the estimation of the number of motors involved in the transport process. Above 
all, we believe that our study will stimulate further interest in understanding the statistical properties of bidirectional cargo motion. 

\acknowledgments
DB acknowledges Dibyendu Das and Ambarish Kunwar for useful discussions. The authors thank Roop Mallik for helpful comments on an earlier version of 
the manuscript. The authors also thank an anonymous referee for drawing their attention to the technique discussed in Appendix A.

\appendix
\section{Derivation of dressed propagators by the method of `splitting probabilities'}

The calculation of direction reversal or direction preserving probabilities ($\phi(\gamma,\beta | \alpha)$) in this paper utilizes the concept of 
dressed propagators $G(\beta|\alpha)$, which we had constructed using the more fundamental bare propagators $\Phi(\beta|\alpha)$. The dressed 
propagator $G(\beta | \alpha)$ includes infinite cyclic transitions between degenerate $\alpha$ states ($\{\alpha_i\}$) of the cargo, for given 
initial and final states. In the present formalism, the modification of the bare propagators by these cyclic transitions involved summing a geometric 
series. However, this method is not the unique; $G(\beta|\alpha)$ can be also determined using a somewhat different method called `method of splitting 
probabilities' \cite{Kampen}. 

Let us consider a stochastic process with more than one absorbing state. The splitting probability for a certain absorbing state is defined as the 
probability that the system reaches it before reaching the others. In other words, it is the transition probability to one of its absorbing states. 
As the system has to be absorbed in one of these states in the long-time limit, the sum of all splitting probabilities is equal to unity. 

The bare propagators $\Phi(\beta | \alpha)$, defined in $(D=1,K=1)$ and $(D=2,K=1)$ cases are indeed splitting probabilities as they are the 
transition probabilities between a starting state $\alpha$ and a final state $\beta$, the latter being treated temporarily as an absorbing state. The 
dressed propagators $G(\beta | \alpha)$ defined in $(D=2,K=1)$ case are higher order splitting probabilities in this sense, and can be constructed 
out of the bare propagators.  It is known that the splitting probabilities follow certain identity relations\cite{Kampen}, which are specific to each 
problem. Here, we exploit this feature to determine the dressed propagators $G(\beta | \alpha)$ for the case $(D=2,K=1)$.

Let, $\{\alpha_i\}=\{m_i\}$ or $\{\tau_i\}$, and $\{\beta_i\}=\{m_i\}$ or $\{\tau_i\}$ such that $\{\alpha_i\} \neq \{\beta_i\}$. Then, the splitting 
probabilities $G(\beta_j|\alpha_i)$ and $G(\beta_j|\alpha_j)$ ($i,j=1,2$) can be easily seen to satisfy the identities

\begin{eqnarray}
 G(\beta_j|\alpha_i)=\Phi(\alpha_j|\alpha_i)G(\beta_j|\alpha_j ) ~~~i\neq j,\\ 
G(\beta_j|\alpha_j)=\Phi(\beta_j|\alpha_j)+\Phi(\alpha_i|\alpha_j) G(\beta_j|\alpha_i),
\label{eq:eq18}
\end{eqnarray}

solving which, it follows that

\begin{eqnarray}
 G(\beta_j|\alpha_i)=\frac{\Phi(\alpha_j|\alpha_i)\Phi(\beta_j|\alpha_j)}{1-\Phi(\alpha_j|\alpha_i)\Phi(\alpha_i|\alpha_j)}  ~~i\neq j,\\
G(\beta_j|\alpha_j)=\frac{\Phi(\beta_j|\alpha_j)}{1-\Phi(\alpha_j|\alpha_i)\Phi(\alpha_i|\alpha_j)}.
\label{eq:eq19}
\end{eqnarray}

which are seen to be identical to the relations in Eq.\ref{eq:3++}, for all combinations of initial and final states.

\section{Relation between Memory and Bias parameters}

Starting from the definition in Eq.\ref{eq:eq1++}, for a $(D=2,K=1)$ system, we have $\nu=\gamma_{p_1}-\sum_\ell \gamma_{m_\ell}$ ($\ell=1,2$). Using the 
parameters $\eta_\ell$ defined in Sec. III A, we arrive at the following relation:

\begin{equation}
\nu= \frac{1-\sum_{\ell} \eta_\ell}{1+\sum_{\ell} \eta_\ell}.
\label{eq:eq23}
\end{equation}

From Eq.\ref{eq:eq7}, we have 

\begin{equation}
\sum_{\ell}\eta_{\ell}=\sum_{i,j,\ell} \eta_i  G(m_\ell,\tau_j|m_i)+  \sum_{\ell} G(m_{\ell},\tau_1|p_1).
\label{eq:eq24c}
\end{equation}

The following normalization conditions clearly apply:

\begin{widetext}
\begin{eqnarray}
 G(p_1,\tau_1|p_1)+ \sum_{\ell} G(m_{\ell},\tau_1|p_1)+ G(o|p_1)=1,  \label{eq:eq21a}\\
\sum_{j,\ell} G(m_\ell,\tau_j|m_i)+ \sum_{j} G(p_1,\tau_j|m_i) + G(o|m_i) =1 ,
\label{eq:eq21b}
\end{eqnarray}
\end{widetext}

where the terms $G(o|p_1)$ and $G(o|m_i)$ give the probability of complete detachment of the cargo, from initial states $p_1$ and $m_i$ respectively. 

Let us now assume $G(o|p_1),G(o|m_i)\ll 1$, and use the identity $G(p_1,\tau_1|p_1)\equiv \phi(p,\tau|p)$ in Eq.\ref{eq:eq21a}, which leads to 
$\sum_{\ell} G(m_{\ell},\tau_1|p_1)\simeq 1-\phi(p,\tau|p)$. Finally, using Eq.\ref{eq:eq21b} in Eq.\ref{eq:eq6b} and using the relation 
$\phi(m,\tau|m)=1-\phi(p,\tau|m)$ leads to a second relation $\sum_{i,j,\ell} \eta_i  G(m_\ell,\tau_j|m_i)\simeq \phi(m,\tau|m)\sum_{\ell}\eta_{\ell}$. 
Using these approximate relations in Eq.\ref{eq:eq24c}, we arrive at the equation

\begin{equation}
\sum_{\ell}\eta_{\ell}\simeq \frac{1-\phi(p,\tau|p)}{1-\phi(m,\tau|m)},
\label{eq:eq22}
\end{equation}

the substitution of which in Eq.\ref{eq:eq23} leads to the expression in Eq.\ref{eq:eq12}, after realizing that $\phi(p,\tau|p)+\phi(m,\tau|m)=1+\mu$ 
and using the definition of the asymmetry coefficient $\eta$ in Eq.\ref{eq:eq1+}. Further, using the relation $\gamma_{p_1}\sum_{\ell}\eta_{\ell}=\sum_{\ell} \gamma_{m_{\ell}}$ along with the condition of normalization ($\gamma_{p_1}+\sum_{\ell} \gamma_{m_{\ell}}=1$) in Eq.\ref{eq:eq22}, it can also be shown that, 

\begin{eqnarray}
\gamma_{p_1}\simeq \frac{\phi(p,\tau|m)}{1-\mu} ~~;~~ \sum_\ell \gamma_{m_\ell}\simeq \frac{\phi(m,\tau|p)}{1-\mu}.
\label{eq:eq22abc}
\end{eqnarray}

\section{Proof that $C(1)=\mu$ under the `no detachment approximation'}

In steady state conditions, $\langle S_i\rangle=\gamma_{p}-\gamma_m=\nu$, the bias parameter. Therefore, Eq.\ref{eq:eq13} becomes 

\begin{equation}
 C(n)=\frac{\langle S_i S_{i+n}\rangle-\nu^2}{1-\nu^2} ~;~ n\geq0.
\label{eq:eq24}
\end{equation}

For $n=1$, it follows from the definitions of the propagators that 

\begin{widetext}
\begin{eqnarray}
\langle S_iS_{i+1}\rangle=\sum_{i,j,\ell} \gamma_{m_i} G(m_{\ell},\tau_{j}|m_i)- \sum_{i,j} \gamma_{m_i} G(p_1,\tau_{j}|m_i) + \gamma_{p_1} G(p_{1},\tau_{1}|p_1)- \gamma_{p_1} \sum_{\ell} G(m_{\ell},\tau_{j}|p_1).
\label{eq:eq25}
\end{eqnarray}
\end{widetext}

If we now assume that the probability of complete detachment of the cargo is small, i.e.,  $G(o|p_1),G(o|m_i)\ll 1$, then from Eq.\ref{eq:eq21a} and 
Eq.\ref{eq:eq6a} we can write  $G(p_{1},\tau_{1}|p_1)\simeq \phi(p,\tau|p)$ and  $\sum_{\ell} G(m_{\ell},\tau_{1}|p_1)\simeq \phi(m,\tau|p)$. 
Similarly, using Eq.\ref{eq:eq21b} and Eq.\ref{eq:eq6b} we can write $\sum_{i,j,\ell} \gamma_{m_i}  G(m_\ell,\tau_j|m_i)\simeq \phi(m,\tau|m)\sum_{i}\gamma_{m_i}$ 
and  $\sum_{i,j,\ell} \gamma_{m_i}  G(p_1,\tau_j|m_i)\simeq \phi(p,\tau|m)\sum_{i}\gamma_{m_i}$. Therefore, Eq.\ref{eq:eq25} becomes

\begin{widetext}
\begin{eqnarray}
\langle S_iS_{i+1}\rangle= [\phi(p,\tau|p)-\phi(m,\tau|p)]\gamma_{p_1} +  [\phi(m,\tau|m)-\phi(p,\tau|m)] \sum_{\ell} \gamma_{m_{\ell}}.
\label{eq:eq26}
\end{eqnarray}
\end{widetext}

Noting that $\phi(m,\tau|p)=\phi(m,\tau|m)-\mu$ and $\phi(p,\tau|m)=\phi(p,\tau|p)-\mu$, Eq.\ref{eq:eq26} can be rewritten as

\begin{equation}
\langle S_iS_{i+1}\rangle=\mu + \eta \nu,
\label{eq:eq28}
\end{equation}

where we have used the fact that $\gamma_{p_1}+\sum_{\ell} \gamma_{m_{\ell}}=1$, $\nu=\gamma_{p_1}-\sum_{\ell} \gamma_{m_{\ell}}$ 
and also from Eq.\ref{eq:eq1+}, $\eta=\phi(p,\tau|p)-\phi(m,\tau|m)$. Now, using Eq.\ref{eq:eq28} in Eq.\ref{eq:eq24}, we find 

\begin{equation}
 C(1)= \frac{\mu + \eta \nu-\nu^2}{1-\nu^2}.
\label{eq:eq29}
\end{equation}

Finally, substituting Eq.\ref{eq:eq12} in Eq.\ref{eq:eq29}, we arrive at the simple and elegant result $C(1)=\mu$.


\begin{thebibliography}{30}
\bibitem{Astumian} R. D. Astumian and M. Bier, Phys. Rev. Lett. {\bf 72} 1766 (1994).
\bibitem{Magnasco1} M. O. Magnasco, Phys. Rev. Lett. {\bf 71} 1477 (1993); {\it ibid} {\bf 72} 2656 (1994).
\bibitem{Prost} J. Prost, J. F. Chauwin, L. Peliti and A. Ajdari, Phys. Rev. Lett. {\bf 72} 2652 (1994).
\bibitem{Astumian2} R. D. Astumian, Science {\bf 276} 917 (1997).
\bibitem{Julicher} F. Julicher, A. Ajdari and J. Prost, Rev. Mod. Phys. {\bf 69} 1269 (1997).
\bibitem{Fisher2} M. E. Fisher and A. B. Kolomeisky,   Proc. Natl. Acad. Sci. USA {\bf 96} 6597 (1999),  {\it ibid} {\bf 98} 7748 (2001).
\bibitem{Parmeggiani}A. Parmeggiani, F. Julicher, L. Peliti and J. Prost, Europhys. Lett. {\bf 56} 603 (2001). 
\bibitem{Fisher} M. E. Fisher and Y. C. Kim,  Proc. Natl. Acad. Sci. USA {\bf 102} 16209 (2005).
\bibitem{Chowdhury} K. Nishinari, Y. Okada, A. Schadschneider and D. Chowdhury, Phys. Rev. Lett. {\bf 95} 118101 (2005).
\bibitem{Fisher3} A. B. Kolomeisky and M. E. Fisher, Annu. Rev. Phys. Chem. {\bf 58} 675 (2007).
\bibitem{Zhang} Y. Zhang, Phys. Rev. E {\bf 84} 031104 (2011).
\bibitem{Debashish} D. Chowdhury, Phys. Rep. {\bf 529} 1 (2013).
\bibitem{Badoual} M. Badoual, F. J\"ulicher and J. Prost, Proc. Natl. Acad. Sci. USA {\bf 99} 6696 (2002).
\bibitem{Klumpp} S. Klumpp and R. Lipowsky,  Proc. Natl. Acad. Sci. USA {\bf 102}  17284 (2005).
\bibitem{SKlumpp} M. J. I. M\"uller, S. Klumpp and R. Lipowsky,  J. Stat. Phys. {\bf 133} 1059 (2008).
\bibitem{Muller} M. J. I. M\"uller , S. Klumpp and R. Lipowsky, Proc. Natl. Acad. Sci. USA {\bf 105} 4609 (2008).
\bibitem{Mjmuller} M. J. I. M\"uller, S. Klumpp and R. Lipowsky,  Biophys. J. {\bf 98} 2610 (2010).
\bibitem{Gross} S. P. Gross, Phys. Biol. {\bf 1} R1 (2004).
\bibitem{Welte} M. A. Welte, Curr. Biol. {\bf 14} R525 (2004).
\bibitem{Deepak} D. Bhat and M. Gopalakrishnan, Phys. Biol. {\bf 9} 046003 (2012).
\bibitem{Maly} I. V. Maly and I. A. Vorobjev, Cell Biol. Int. {\bf 26} 791 (2002). 
\bibitem{Leidel} C. Leidel, R. A. Longoria, F. M. Gutierrez, and G. T. Shubeita, Biophys. J. {\bf 103} 492 (2012).
\bibitem{Yunxin} Y. Zhang, Phys. Rev. E {\bf 87} 052705 (2013).
\bibitem{schnitzer} M. J. Schnitzer, Phys. Rev. E {\bf 48} 2553 (1993).
\bibitem{sambeth} R. Sambeth and A. Baumgaertner, Phys. Rev. Lett. {\bf 86} 5196 (2001).
\bibitem{Kunwar} A. Kunwar {\it et al},  Proc. Natl. Acad. Sci. USA {\bf 108} 18960 (2011).
\bibitem{phagosome} A. K. Rai, A. Rai, A. J. Ramaiya, R. Jha, and R. Mallik, Cell {\bf 152} 1 (2013).
\bibitem{Goldstein} S. Goldstein, Quart. J. Mech. Appl. Math. {\bf 4} 129 (1951).
\bibitem{supp} See Supplemental Material at [URL will be inserted by publisher] for calculations of waiting time distribution as well as simulation results for velocity autocorrelation function.
\bibitem{Ronojoy} S. Saha, S. Ghose, R. Adhikari, and A. Dua, Phys. Rev. Lett. {\bf 107} 218301 (2011).
\bibitem{Gillespie} D. T. Gillespie, J. Phys. Chem. {\bf 81} 2340 (1977).
\bibitem{Soppina} V. Soppina, A. K. Rai, A. J. Ramaiya, P. Barak, and R. Mallik, Proc. Natl. Acad. Sci. USA {\bf 106} 19381 (2009).
\bibitem{KSvoboda} K. Svoboda and S. M. Block, Cell {\bf 77} 773 (1994).
\bibitem{RDVale} R. D. Vale, T. S. Funatsu, D. W. Pierce, L. Romberg, Y. Harada and T. Yanagida, Nature {\bf 380} 451 (1996).
\bibitem{Coppin} C. M. Coppin, D. W. Pierce, L. Hsu and R. D. Vale, Proc. Natl. Acad. Sci. USA {\bf 94} 8539 (1997).
\bibitem{MAWelte} M. A. Welte, S. P. Gross, M. Postner, S. M. Block and E. F. Wieschaus, Cell {\bf 92} 547 (1998).
\bibitem{MJSchnitzer} M. J. Schnitzer, K. Visscher and S. M. Block, Nat. Cell. Biol. {\bf 2} 718 (2000).
\bibitem{SToba} S. Toba, T. M. Watanabe, L. Yamaguchi-Okimoto, Y. Y. Toyoshima and H. Higuchi, Proc. Natl. Acad. Sci. USA {\bf 103} 5741 (2006).
\bibitem{Vaishnavi} V. Ananthanarayanan, M. Schattat, S. K. Vogel, A. Krull, N. Pavin, and I. M. Toli\`c-N\o rrelykke, Cell {\bf 153} 1526 (2013).
\bibitem{DAE} D. Bhat and M. Gopalakrishnan, AIP Conf. Proc. {\bf 1512}, 140 (2012).
\bibitem{remark} An analogy may be made with the spin correlation function in the 1D Ising model, which decays exponentially with the number of sites, see e.g., P.M. Chaikin and T.C. Lubensky, {\it Principles of condensed matter physics} (Cambridge University Press, 1995).
\bibitem{codling} E. A. Codling, M. J. Plank and S. Benhamou, J. R. Soc. Interface {\bf 5} 813 (2008).
\bibitem{furth} R. F\"urth, Schwankungserscheinungen in der Physik, Sammlung Vieweg, Braunschweig (1920).
\bibitem{taylor} G.I. Taylor, Proc. London Math. Soc. {\bf 20} 196 (1921/22).
\bibitem{patlak} C. S. Patlak, Bull. Math. Biophys. {\bf 15} 311 (1953).
\bibitem{weiss} G. H. Weiss, Physica A {\bf 311} 381 (2002).
\bibitem{Ricardo} R. Garcia-Pelayo, Physica A {\bf 384} 143 (2007).
\bibitem{hermann} S. Hermann and P. Vallois, e-print arXiv:0810.0650v1 [math.PR] (2008).
\bibitem{Hendricks} A. G. Hendricks, E. Perlson, J. L. Ross, H. W. Schroeder III, M. Tokito, and E. L. F. Holzbaur, Curr. Biol. {\bf 20} 697 (2010).
\bibitem{Kampen} N. G. van Kampen, {\it Stochastic Processes in Physics and Chemistry} (Elsevier Science Publishers B.V., North-Holland, Amsterdam, 1997), Chap. XII. 
\end{thebibliography}
\end{document}